%% file: qapl2012.tex
\documentclass[copyright]{eptcs}

\providecommand{\event}{QAPL 2012}

\providecommand{\firstpage}{1}

\usepackage{amssymb}
\usepackage{epsf}
\input{Commands}

\input{Environments}

\title{Weak Markovian Bisimulation Congruences and Exact CTMC-Level Aggregations for Concurrent Processes}

\author{Marco Bernardo
\institute{Dipartimento di Scienze di Base e Fondamenti -- Universit\`a di Urbino -- Italy}}

\def\titlerunning{Weak Markovian Bisimulation Congruences and Exact CTMC-Level Aggregations}
\def\authorrunning{M.\ Bernardo}

\begin{document}

\maketitle

\begin{abstract}

\noindent
We have recently defined a weak Markovian bisimulation equivalence in an integrated-time setting, which
reduces sequences of exponentially timed internal actions to individual exponentially timed internal actions
having the same average duration and execution probability as the corresponding sequences. This weak
Markovian bisimulation equivalence is a congruence for sequential processes with abstraction and turns out
to induce an exact CTMC-level aggregation at steady state for all the considered processes. However, it is
not a congruence with respect to parallel composition. In this paper, we show how to generalize the
equivalence in a way that a reasonable tradeoff among abstraction, compositionality, and exactness is
achieved for concurrent processes. We will see that, by enhancing the abstraction capability in the presence
of concurrent computations, it is possible to retrieve the congruence property with respect to parallel
composition, with the resulting CTMC-level aggregation being exact at steady state only for a certain subset
of the considered processes.

\end{abstract}

%
%
\section{Introduction}\label{intro}
%
%

Several Markovian behavioral equivalences (see~\cite{ABC} and the references therein) have been proposed in
the literature for relating and manipulating system models with an underlying continuous-time Markov chain
(CTMC)~\cite{Ste} semantics. However, only a few of them are provided with the useful capability of
abstracting from internal actions. In particular, \cite{Ber} has recently addressed the case in which
internal actions are exponentially timed -- rather than immediate like in~\cite{Her} -- by defining a weak
Markovian bisimulation equivalence inspired by the weak (Markovian) isomorphism of~\cite{Hil}. The idea is
to reduce to \textit{individual} exponentially timed internal transitions all the \textit{sequences} of
exponentially timed internal transitions that traverse states enabling \textit{only} exponentially timed
internal actions, with the reduction preserving the average duration and the execution probability of the
original sequences.

From a stochastic viewpoint, this reduction amounts to replacing hypoexponentially distributed durations
with exponentially distributed durations having the same expected value. As a consequence, processes related
by the weak Markovian bisimulation equivalence of~\cite{Ber} may not possess the same transient performance
measures, unless they refer to properties of the form mean time to certain events. However, those processes
certainly possess the same steady-state performance measures, because the aggregation induced by the
considered equivalence on the CTMC underlying each process has been shown to be exact at steady state.

The weak Markovian bisimulation equivalence of~\cite{Ber} is not a congruence with respect to parallel
composition, a fact that limits its usefulness for compositional state space reduction purposes. The
contribution of this paper is to show that compositionality can be retrieved by enhancing the abstraction
capability of the considered equivalence in the presence of parallel composition. The basic idea is allowing
a sequence of exponentially timed internal transitions originated from a sequential process to be reduced
also in the case in which that process is composed in parallel with other processes enabling
\textit{observable} actions. Unfortunately, there is a price to pay for achieving compositionality:
exactness at steady state will no longer hold for all processes, but only for processes with no
synchronization at all and processes whose synchronizations do not take place right before the sequences to
be reduced.

This paper is organized as follows. After introducing a Markovian process calculus in Sect.~\ref{mpc} and
recalling strong and weak Markovian bisimilarity in Sect.~\ref{strongweak}, in Sect.~\ref{concproc} we
develop a variant of weak Markovian bisimilarity that deals with parallel composition and we investigate its
congruence and exactness properties. Finally, in Sect.~\ref{concl} we provide some concluding remarks.

%
%
\section{Concurrent Markovian Processes}\label{mpc}
%
%

In order to study properties such as congruence of the variant (to be defined) of the weak Markovian
bisimilarity of~\cite{Ber}, we introduce typical behavioral operators through a Markovian process calculus
\linebreak (MPC for short). In~\cite{Ber}, we have considered sequential processes with abstraction built
from operators like the inactive process, exponentially timed action prefix, alternative composition,
recursion, and hiding. Here, we include parallel composition too, so as to be able to represent concurrent
processes.

As usual, we denote the internal action by $\tau$ and we assume that the resulting concurrent processes are
governed by the race policy: if several exponentially timed actions are simultaneously enabled, the action
that is executed is the one sampling the least duration. We also assume that the duration of an action
deriving from the synchronization of two exponentially timed actions is exponentially distributed with a
rate obtained by applying (like, e.g., in~\cite{HR}) some commutative and associative operation denoted by
$\otimes$ to the rates of the two original actions.

	\begin{definition}

Let $\ms{Act}_{\rm M} = \ms{Name} \times \realns_{> 0}$ be a set of actions, where $\ms{Name} =
\ms{Name}_{\rm v} \cup \{ \tau \}$ is a set of action names -- ranged over by $a, b$ -- and $\realns_{> 0}$
is a set of action rates -- ranged over by $\lambda, \mu, \gamma$. Let $\ms{Var}$ be a set of process
variables ranged over by $X, Y$. The process language $\calpl_{\rm M}$ is generated by the following syntax:
\[\begin{array}{|rcll|}
\hline
P & \; ::= \; & \nil & \hspace{0.5cm} \textrm{inactive process} \\
& | & \lap a, \lambda \rap . P & \hspace{0.5cm} \textrm{exponentially timed action prefix} \\
& | & P + P & \hspace{0.5cm} \textrm{alternative composition} \\
& | & X & \hspace{0.5cm} \textrm{process variable} \\
& | & {\rm rec} \, X : P & \hspace{0.5cm} \textrm{recursion} \\
& | & P / H & \hspace{0.5cm} \textrm{hiding} \\
& | & P \pco{S} P & \hspace{0.5cm} \textrm{parallel composition} \\
\hline
\end{array}\]
where $a \in \ms{Name}$, $\lambda \in \realns_{> 0}$, $X \in \ms{Var}$, and $H, S \subseteq \ms{Name}_{\rm
v}$. We denote by $\procs_{\rm M}$ the set of closed and guarded process terms of $\calpl_{\rm M}$ -- ranged
over by $P, Q$.
\fullbox

	\end{definition}

In order to distinguish between process terms such as $\lap a, \lambda \rap . \nil + \lap a, \lambda \rap .
\nil$ and $\lap a, \lambda \rap . \nil$, like in~\cite{Ber} the semantic model $\lsp P \rsp_{\rm M}$ for a
process term $P \in \procs_{\rm M}$ is a labeled multitransition system that takes into account the
multiplicity of each transition, intended as the number of different proofs for the transition derivation.
The multitransition relation of $\lsp P \rsp_{\rm M}$ is contained in the smallest multiset of elements of
$\procs_{\rm M} \times \ms{Act}_{\rm M} \times \procs_{\rm M}$ that satisfies the operational semantic rules
in Table~\ref{mpcsos} -- where $\{ \_ \hookrightarrow \_ \}$ denotes syntactical replacement -- and keeps
track of all the possible ways of deriving each of its transitions.

	\begin{table}[thb]

\[\begin{array}{|c|}
\hline
\hspace*{1.4cm} (\textsc{Pre}_{\rm M}) \hspace{0.2cm} {\infr{}{\lap a, \lambda \rap . P \arrow{a,
\lambda}{\rm M} P}} \hspace{0.8cm}
(\textsc{Rec}_{\rm M}) \hspace{0.2cm} {\infr{P \{ \textrm{rec} \, X : P \hookrightarrow X \} \arrow{a,
\lambda}{\rm M} P'}{\textrm{rec} \, X : P \arrow{a, \lambda}{\rm M} P'}} \\[0.8cm]
(\textsc{Alt}_{\rm M, 1}) \hspace{0.2cm} {\infr{P_{1} \arrow{a, \lambda}{\rm M} P'}{P_{1} + P_{2} \arrow{a,
\lambda}{\rm M} P'}} \hspace{0.8cm}
(\textsc{Alt}_{\rm M, 2}) \hspace{0.2cm} {\infr{P_{2} \arrow{a, \lambda}{\rm M} P'}{P_{1} + P_{2} \arrow{a,
\lambda}{\rm M} P'}} \\[0.8cm]
(\textsc{Hid}_{\rm M, 1}) \hspace{0.2cm} {\infr{P \arrow{a, \lambda}{\rm M} P' \hspace{0.5cm} a \notin H}{P
/ H \arrow{a, \lambda}{\rm M} P' / H}} \hspace{0.8cm}
(\textsc{Hid}_{\rm M, 2}) \hspace{0.2cm} {\infr{P \arrow{a, \lambda}{\rm M} P' \hspace{0.5cm} a \in H}{P / H
\arrow{\tau, \lambda}{\rm M} P' / H}} \\[0.8cm]
(\textsc{Par}_{\rm M, 1}) \hspace{0.2cm} {\infr{P_{1} \arrow{a, \lambda}{\rm M} P'_{1} \hspace{0.5cm} a
\notin S}{P_{1} \pco{S} P_{2} \arrow{a, \lambda}{\rm M} P'_{1} \pco{S} P_{2}}} \hspace{0.8cm}
(\textsc{Par}_{\rm M, 2}) \hspace{0.2cm} {\infr{P_{2} \arrow{a, \lambda}{\rm M} P'_{2} \hspace{0.5cm} a
\notin S}{P_{1} \pco{S} P_{2} \arrow{a, \lambda}{\rm M} P_{1} \pco{S} P'_{2}}} \\[0.8cm]
(\textsc{Syn}_{\rm M}) \hspace{0.2cm} {\infr{P_{1} \arrow{a, \lambda_{1}}{\rm M} P'_{1} \hspace{0.5cm} P_{2}
\arrow{a, \lambda_{2}}{\rm M} P'_{2} \hspace{0.5cm} a \in S}{P_{1} \pco{S} P_{2} \arrow{a, \lambda_{1}
\otimes \lambda_{2}}{\rm M} P'_{1} \pco{S} P'_{2}}} \\
\hline
\end{array}\]

\caption{Structured operational semantic rules for MPC}
\label{mpcsos}

	\end{table}

%
%
\section{Strong and Weak Markovian Bisimulation Equivalences}\label{strongweak}
%
%

The notion of strong bisimilarity for MPC is based on the comparison of exit rates~\cite{Hil,HR}. The exit
rate of a process term $P \in \procs_{\rm M}$ with respect to action name $a \in \ms{Name}$ and destination
$D \subseteq \procs_{\rm M}$ is the rate at which $P$ can execute actions of name $a$ that lead to $D$:
\cws{0}{\ms{rate}(P, a, D) \: = \: \sum \lmp \lambda \in \realns_{> 0} \mid \exists P' \in D \ldotp P
\arrow{a, \lambda}{\rm M} P' \rmp}
where $\lmp$ and $\rmp$ are multiset delimiters and the summation is taken to be zero if its multiset is
empty. By summing up the rates of all the actions of $P$, we obtain the total exit rate of $P$, i.e.,
$\ms{rate}_{\rm t}(P) = \sum_{a \in \ms{Name}} \ms{rate}(P, a, \procs_{\rm M})$, which is the reciprocal of
the average sojourn time associated with $P$.

	\begin{definition}

An equivalence relation $\calb$ over $\procs_{\rm M}$ is a Markovian bisimulation iff, whenever $(P_{1},
P_{2}) \in \calb$, then for all action names $a \in \ms{Name}$ and equivalence classes $D \in \procs_{\rm M}
/ \calb$:
\cws{0}{\ms{rate}(P_{1}, a, D) \: = \: \ms{rate}(P_{2}, a, D)}
Markovian bisimilarity $\sbis{\rm MB}$ is the largest Markovian bisimulation.
\fullbox

	\end{definition}

\noindent
As shown in~\cite{Hil,HR,Buc,DHS}, the relation $\sbis{\rm MB}$ possesses the following properties:

	\begin{itemize}

\item $\sbis{\rm MB}$ is a congruence with respect to all the operators of MPC as well as recursion.

\item $\sbis{\rm MB}$ has a sound and complete axiomatization whose basic laws are shown below:
\[\begin{array}{|lrcl|}
\hline
(\cala_{{\rm MB}, 1}) \quad & P_{1} + P_{2} & = & P_{2} + P_{1} \\
(\cala_{{\rm MB}, 2}) \quad & (P_{1} + P_{2}) + P_{3} & = & P_{1} + (P_{2} + P_{3}) \\
(\cala_{{\rm MB}, 3}) \quad & P + \nil & = & P \\
(\cala_{{\rm MB}, 4}) \quad & \lap a, \lambda_{1} \rap . P + \lap a, \lambda_{2} \rap . P & = & \lap a,
\lambda_{1} + \lambda_{2} \rap . P \\
\hline
\end{array}\]
The last one encodes the race policy and hence replaces the idempotency law $P + P = P$ valid for
nondeterministic processes. The other laws are the usual distribution laws for the hiding operator and the
expansion law for the parallel composition operator.

\item $\sbis{\rm MB}$ induces a CTMC-level aggregation known as ordinary lumpability, which is exact both at
steady state and at transient state.

\item $\sbis{\rm MB}$ can be decided in polynomial time for all finite-state processes.

	\end{itemize}

In~\cite{Ber}, we have weakened the distinguishing power of $\sbis{\rm MB}$ by relating sequences of
exponentially timed $\tau$-actions to single exponentially timed $\tau$-actions having the same average
duration and execution probability as the sequences. Given $P \in \procs_{\rm M}$, we say that $P$ is stable
if $P \hspace{0.3cm}\not\hspace{-0.5cm}\arrow{\tau, \lambda}{\rm M} P'$ for all $\lambda$ and $P'$,
otherwise we say that it is unstable. In the latter case, we say that $P$ is fully unstable iff, whenever $P
\arrow{a, \lambda}{\rm M} P'$, then $a = \tau$. We denote by $\procs_{\rm M, fu}$ and $\procs_{\rm M, nfu}$
the sets of process terms of $\procs_{\rm M}$ that are fully unstable and not fully unstable, respectively.

The most natural candidates as sequences of exponentially timed $\tau$-actions to abstract are those
labeling computations that traverse fully unstable states.

	\begin{definition}\label{reducompdef}

Let $n \in \natns_{> 0}$ and $P_{1}, P_{2}, \dots, P_{n + 1} \in \procs_{\rm M}$. A computation $c$ of
length $n$ from $P_{1}$ to $P_{n + 1}$ having the form $P_{1} \arrow{\tau, \lambda_{1}}{\rm M} P_{2}
\arrow{\tau, \lambda_{2}}{\rm M} \dots \arrow{\tau, \lambda_{n}}{\rm M} P_{n + 1}$ is reducible iff $P_{i}
\in \procs_{\rm M, fu}$ for all $i = 1, \dots, n$.
\fullbox

	\end{definition}

\noindent
If reducible, the computation $c$ above can be reduced to a single exponentially timed $\tau$-transition
whose rate is obtained from the positive real value below:
\cws{0}{\ms{probtime}(c) \: = \: \left( \prod\limits_{i = 1}^{n} {\lambda_{i} \over \ms{rate}(P_{i}, \tau,
\procs_{\rm M})} \right) \cdot \left( \sum\limits_{i = 1}^{n} {1 \over \ms{rate}(P_{i}, \tau, \procs_{\rm
M})} \right)}
by leaving its first factor unchanged and taking the reciprocal of the second one. The value
$\ms{probtime}(c)$ is a measure of the execution probability of $c$ (first factor: product of the execution
probabilities of the transitions of~$c$) and the average duration of $c$ (second factor: sum of the average
sojourn times in the states traversed by~$c$).

The weak variant of $\sbis{\rm MB}$ defined in~\cite{Ber} is such that (i) processes in $\procs_{\rm M,
nfu}$ are dealt with as in $\sbis{\rm MB}$ and (ii) the length of reducible computations from processes in
$\procs_{\rm M, fu}$ to processes in $\procs_{\rm M, nfu}$ is abstracted away while preserving the execution
probability and the average duration of those computations. In the latter case, we need to lift measure
$\ms{probtime}$ from individual reducible computations to multisets of reducible computations. Denoting by
$\ms{reducomp}(P, D, t)$ the multiset of reducible computations from $P \in \procs_{\rm M, fu}$ to some $P'$
in $D \subseteq \procs_{\rm M}$ whose average duration is $t \in \realns_{> 0}$, we consider the following
$t$-indexed multiset of sums of $\ms{probtime}$ measures:
\cws{8}{\ms{pbtm}(P, D) \: = \: \bigcup\limits_{t \in \realns_{> 0} \: {\rm s.t.} \: \ms{reducomp}(P, D, t)
\neq \emptyset} \hspace{0.5cm} \lmp \sum\limits_{c \in \ms{reducomp}(P, D, t)} \hspace{-0.5cm}
\ms{probtime}(c) \rmp}

	\begin{definition}\label{wmbedef}

An equivalence relation $\calb \subseteq (\procs_{\rm M, nfu} \times \procs_{\rm M, nfu}) \cup (\procs_{\rm
M, fu} \times \procs_{\rm M, fu})$ is a weak Markovian bisimulation iff for all $(P_{1}, P_{2}) \in \calb$:

		\begin{itemize}

\item If $P_{1}, P_{2} \in \procs_{\rm M, nfu}$, then for all $a \in \ms{Name}$ and equivalence classes $D
\in \procs_{\rm M} / \calb$:
\cws{10}{\hspace*{-0.6cm} \ms{rate}(P_{1}, a, D) \: = \: \ms{rate}(P_{2}, a, D)}

\item If $P_{1}, P_{2} \in \procs_{\rm M, fu}$, then for all equivalence classes $D \in \procs_{\rm M,
nfu} / \calb$:
\cws{12}{\hspace*{-0.6cm} \ms{pbtm}(P_{1}, D) \: = \: \ms{pbtm}(P_{2}, D)}

		\end{itemize}

\noindent
Weak Markovian bisimilarity $\wbis{\rm MB}$ is the largest weak Markovian bisimulation.
\fullbox

	\end{definition}

	\begin{example}\label{wmbeex}

Typical cases of weakly Markovian bisimilar process terms are:
\cws{0}{\lap \tau, \mu \rap . \lap \tau, \gamma \rap . Q \hspace{0.5cm} \lap \tau, \gamma \rap . \lap \tau,
\mu \rap . Q \hspace{0.5cm} \lap \tau, {\mu \cdot \gamma \over \mu + \gamma} \rap . Q}
and:
\cws{0}{\begin{array}{l}
\lap \tau, \mu \rap . (\lap \tau, \gamma_{1} \rap . Q_{1} + \lap \tau, \gamma_{2} \rap . Q_{2}) \\
\lap \tau, {\gamma_{1} \over \gamma_{1} + \gamma_{2}} \cdot \left( {1 \over \mu} + {1 \over \gamma_{1} +
\gamma_{2}} \right)^{-1} \hspace{-0.1cm} \rap . Q_{1} + \lap \tau, {\gamma_{2} \over \gamma_{1} +
\gamma_{2}} \cdot \left( {1 \over \mu} + {1 \over \gamma_{1} + \gamma_{2}} \right)^{-1} \hspace{-0.1cm} \rap
. Q_{2} \\
\end{array}}
and:
\cws{0}{\begin{array}{l}
\lap \tau, \mu_{1} \rap . \lap \tau, \gamma \rap . Q_{1} + \lap \tau, \mu_{2} \rap . \lap \tau, \gamma \rap
. Q_{2} \\
\lap \tau, {\mu_{1} \over \mu_{1} + \mu_{2}} \cdot \left( {1 \over \mu_{1} + \mu_{2}} + {1 \over \gamma}
\right)^{-1} \hspace{-0.1cm} \rap . Q_{1} + \lap \tau, {\mu_{2} \over \mu_{1} + \mu_{2}} \cdot \left( {1
\over \mu_{1} + \mu_{2}} + {1 \over \gamma} \right)^{-1} \hspace{-0.1cm} \rap . Q_{2} \\
\end{array}}
where $Q, Q_{1}, Q_{2} \in \procs_{\rm M, nfu}$ (see~\cite{Ber} for the details).
\fullbox

	\end{example}

Similar to weak bisimilarity for nondeterministic processes, $\wbis{\rm MB}$ is not a congruence with
respect to the alternative composition operator. This problem, which has to do with fully unstable process
terms, can be prevented by adopting a construction analogous to the one used in~\cite{Mil} for weak
bisimilarity over nondeterministic process terms. In other words, we have to apply the exit rate equality
check also to fully unstable process terms, with the equivalence classes to consider being the ones with
respect to $\wbis{\rm MB}$.

	\begin{definition}

Let $P_{1}, P_{2} \in \procs_{\rm M}$. We say that $P_{1}$ is weakly Markovian bisimulation congruent to
$P_{2}$, written $P_{1} \obis{\rm MB} P_{2}$, iff for all action names $a \in \ms{Name}$ and equivalence
classes $D \in \procs_{\rm M} / \! \wbis{\rm MB}$:
\cws{10}{\ms{rate}(P_{1}, a, D) \: = \: \ms{rate}(P_{2}, a, D)}
\fullbox

	\end{definition}

\noindent
As shown in~\cite{Ber}, the relation $\obis{\rm MB}$ possesses the following properties:

	\begin{itemize}

\item $\obis{\rm MB}$ is the coarsest congruence -- with respect to all the operators of MPC other than
parallel composition, as well as recursion -- contained in $\wbis{\rm MB}$.

\item $\obis{\rm MB}$ has a sound and complete axiomatization over the set of sequential process terms
(i.e., process terms with no occurrences of the parallel composition operator), whose basic laws are those
of~$\sbis{\rm MB}$ plus the following one (which includes the various cases shown in Ex.~\ref{wmbeex}):
\[\begin{array}{|lrcl|}
\hline
(\cala_{{\rm MB}, 5}) \quad & \lap a, \lambda \rap . \sum\limits_{i \in I} \lap \tau, \mu_{i} \rap .
\sum\limits_{j \in J_{i}} \hspace{-0.1cm} \lap \tau, \gamma_{i, j} \rap . P_{i, j} & = & \\[-0.2cm]
& & & \hspace{-2.1cm} \lap a, \lambda \rap . \sum\limits_{i \in I} \sum\limits_{j \in J_{i}} \hspace{-0.1cm}
\lap \tau, {\mu_{i} \over \mu} \cdot {\gamma_{i, j} \over \gamma} \cdot \left( {1 \over \mu} + {1 \over
\gamma} \right)^{\hspace{-0.1cm} -1} \hspace{-0.2cm} \rap . P_{i, j} \\
\hline
\end{array}\]
where $I \neq \emptyset$ is a finite index set, $J_{i} \neq \emptyset$ is a finite index set for all $i \in
I$, $\mu = \sum_{i \in I} \mu_{i}$, and $\gamma = \sum_{j \in J_{i}} \gamma_{i, j}$ for all $i \in I$.

\item $\obis{\rm MB}$ induces a CTMC-level aggregation called W-lumpability, which is exact only at steady
state and performs reductions consistent with $\cala_{{\rm MB}, 5}$. Moreover, $\obis{\rm MB}$ preserves
transient properties expressed in terms of the mean time to certain events.

\item $\obis{\rm MB}$ can be decided in polynomial time only for those finite-state processes that are not
divergent, i.e., that have no cycles of exponentially timed $\tau$-transitions.

	\end{itemize}

%
%
\section{Compositionality for Concurrent Processes}\label{concproc}
%
%

The relation $\obis{\rm MB}$ is not a congruence with respect to the parallel composition operator, thus
restricting the usefulness for compositional state space reduction purposes of the framework developed
in~\cite{Ber}.

	\begin{example}\label{wmbenocongrparex}

Assuming parallel composition to have lower priority than any other operator, it holds that:
\cws{0}{\lap a, \lambda \rap . \lap \tau, \mu \rap . \lap \tau, \gamma \rap . \nil \: \obis{\rm MB} \: \lap
a, \lambda \rap . \lap \tau, {\mu \cdot \gamma \over \mu + \gamma} \rap . \nil}
while:
\cws{0}{\lap a, \lambda \rap . \lap \tau, \mu \rap . \lap \tau, \gamma \rap . \nil \pco{\emptyset} \lap a',
\lambda' \rap . \nil \: \not\obis{\rm MB} \: \lap a, \lambda \rap . \lap \tau, {\mu \cdot \gamma \over \mu +
\gamma} \rap . \nil \pco{\emptyset} \lap a', \lambda' \rap . \nil}
First of all, we note that:
\cws{0}{\lap \tau, \mu \rap . \lap \tau, \gamma \rap . \nil \pco{\emptyset} \lap a', \lambda' \rap . \nil \:
\not\wbis{\rm MB} \: \lap \tau, {\mu \cdot \gamma \over \mu + \gamma} \rap . \nil \pco{\emptyset} \lap a',
\lambda' \rap . \nil}
In fact, for $a' \neq \tau$ the two process terms are not fully unstable with:
\cws{0}{\begin{array}{rcl}
\ms{rate}(\lap \tau, \mu \rap . \lap \tau, \gamma \rap . \nil \pco{\emptyset} \lap a', \lambda' \rap . \nil,
\tau, [\lap \tau, \gamma \rap . \nil \pco{\emptyset} \lap a', \lambda' \rap . \nil]_{\wbis{\rm MB}}) & = &
\mu \\
\ms{rate}(\lap \tau, {\mu \cdot \gamma \over \mu + \gamma} \rap . \nil \pco{\emptyset} \lap a', \lambda'
\rap . \nil, \tau, [\lap \tau, \gamma \rap . \nil \pco{\emptyset} \lap a', \lambda' \rap . \nil]_{\wbis{\rm
MB}}) & = & 0 \\
\end{array}}
On the other hand, for $a' = \tau$ the two process terms are fully unstable with:
\cws{0}{\begin{array}{rcl}
\ms{pbtm}(\lap \tau, \mu \rap . \lap \tau, \gamma \rap . \nil \pco{\emptyset} \lap a', \lambda' \rap . \nil,
[\nil \pco{\emptyset} \nil]_{\wbis{\rm MB}}) & = & \lmp ({\mu \over \mu + \lambda'} \cdot {\gamma \over
\gamma + \lambda'}) \cdot ({1 \over \mu + \lambda'} + {1 \over \gamma + \lambda'} + {1 \over \lambda'}), \\
& & \hspace*{0.3cm} ({\mu \over \mu + \lambda'} \cdot {\lambda' \over \gamma + \lambda'}) \cdot ({1 \over
\mu + \lambda'} + {1 \over \gamma + \lambda'} + {1 \over \gamma}), \\
& & \hspace*{0.3cm} ({\lambda' \over \mu + \lambda'}) \cdot ({1 \over \mu + \lambda'} + {1 \over \mu} + {1
\over \gamma}) \rmp \\
\ms{pbtm}(\lap \tau, {\mu \cdot \gamma \over \mu + \gamma} \rap . \nil \pco{\emptyset} \lap a', \lambda'
\rap . \nil, [\nil \pco{\emptyset} \nil]_{\wbis{\rm MB}}) & = & \lmp ({{\mu \cdot \gamma \over \mu + \gamma}
\over {\mu \cdot \gamma \over \mu + \gamma} + \lambda'}) \cdot ({1 \over {\mu \cdot \gamma \over \mu +
\gamma} + \lambda'} + {1 \over \lambda'}), \\
& & \hspace*{0.3cm} ({\lambda' \over {\mu \cdot \gamma \over \mu + \gamma} + \lambda'}) \cdot ({1 \over {\mu
\cdot \gamma \over \mu + \gamma} + \lambda'} + {1 \over {\mu \cdot \gamma \over \mu + \gamma}}) \rmp \\
\end{array}}
Thus:
\cws{0}{[\lap \tau, \mu \rap . \lap \tau, \gamma \rap . \nil \pco{\emptyset} \lap a', \lambda' \rap .
\nil]_{\wbis{\rm MB}} \cap [\lap \tau, {\mu \cdot \gamma \over \mu + \gamma} \rap . \nil \pco{\emptyset}
\lap a', \lambda' \rap . \nil]_{\wbis{\rm MB}} \: = \: \emptyset}
and hence:
\cws{0}{\ms{rate}(\lap a, \lambda \rap . \lap \tau, \mu \rap . \lap \tau, \gamma \rap . \nil \pco{\emptyset}
\lap a', \lambda' \rap . \nil, a, [\lap \tau, \mu \rap . \lap \tau, \gamma \rap . \nil \pco{\emptyset} \lap
a', \lambda' \rap . \nil]_{\wbis{\rm MB}}) \: = \: \lambda}
whereas:
\cws{0}{\ms{rate}(\lap a, \lambda \rap . \lap \tau, {\mu \cdot \gamma \over \mu + \gamma} \rap . \nil
\pco{\emptyset} \lap a', \lambda' \rap . \nil, a, [\lap \tau, \mu \rap . \lap \tau, \gamma \rap . \nil
\pco{\emptyset} \lap a', \lambda' \rap . \nil]_{\wbis{\rm MB}}) \: = \: 0}
Also the two divergent process terms $\textrm{rec} \, X : \lap \tau, \mu \rap . \lap \tau, \gamma_{1} \rap .
X$ and $\textrm{rec} \, X : \lap \tau, \mu \rap . \lap \tau, \gamma_{2} \rap . X$, $\gamma_{1} \neq
\gamma_{2}$, are related by $\obis{\rm MB}$ but this no longer holds when placing them in the context $\_
\pco{\emptyset} \lap a', \lambda' \rap . \nil$, $a' \neq \tau$.
\fullbox

	\end{example}

Taking inspiration from the weak isomorphism of~\cite{Hil}, in this section we show how to retrieve full
compositionality by enhancing the abstraction capability of $\obis{\rm MB}$ in the case of concurrent
computations. The price to pay is that exactness will hold at steady state only for a certain class of
processes.

%
\subsection{Revising Weak Markovian Bisimilarity}\label{wmberedef}
%

As we have seen, $\wbis{\rm MB}$ and $\obis{\rm MB}$ abstract from sequences of exponentially timed
$\tau$-actions while preserving (at the computation level) their execution probability and average duration
and (at the system level) \linebreak transient properties expressed in terms of the mean time to certain
events as well as steady-state performance measures. This kind of abstraction has been done in the simplest
possible case: sequences of exponentially timed $\tau$-actions labeling computations that traverse
\textit{fully unstable states}.

In order to achieve compositionality when dealing with concurrent processes, a revision of the notion of
reducible computation is unavoidable. More precisely, we need to address the case of sequences of
exponentially timed $\tau$-actions labeling computations that traverse \textit{unstable states satisfying
certain conditions}. The reason is that, if we view a system description as the parallel composition of
several sequential processes, any of those processes may have local computations traversing \textit{fully
unstable local states}, but in the overall system those local states may be \textit{part of global states
that are not fully unstable}.

For instance, this is the case with the process $\lap \tau, \mu \rap . \lap \tau, \gamma \rap . \nil
\pco{\emptyset} \lap a, \lambda \rap . \nil$, whose underlying labeled multitransition system is depicted
below on the left: \\[0.1cm]
\centerline{\epsfbox{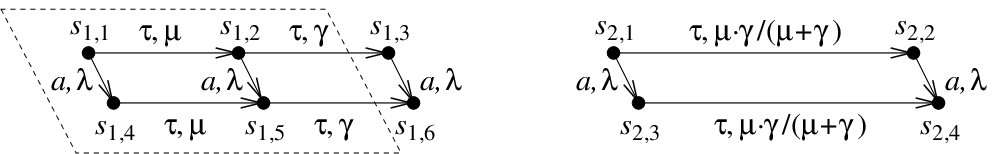}}
As can be noted, the fully unstable local states traversed by the only local computation of the sequential
process $\lap \tau, \mu \rap . \lap \tau, \gamma \rap . \nil$ may become part of unstable global states that
are not fully unstable if $a \neq \tau$. \linebreak Our objective is to change the notion of reducible
computation in such a way that the labeled multitransition system on the left can be regarded as being
weakly Markovian bisimilar to the labeled multitransition system on the right. As can be noted, this implies
that execution probabilities and average durations can only be preserved \textit{at the level of local
computations}, hence transient properties expressed in terms of the mean time to certain events can no
longer be preserved at the system level.

In a concurrent setting, a sequence of exponentially timed $\tau$-actions may be replicated due to
interleaving, in the sense that it may label several computations that share no transition. The revision of
the notion of reducible computation is thus based on the idea that, for each computation that traverses
fully unstable \textit{local} states and is labeled with exponentially timed $\tau$-actions, we have to
recognize -- and take into account at once -- \textit{all the replicas} of that computation and pinpoint
their initial and final states. In our example, there are two replicas with initial states $s_{1, 1}$ and
$s_{1, 4}$ and final states $s_{1, 3}$ and $s_{1, 6}$.

In general, a one-to-one correspondence can be established between the states traversed by any two replicas
by following the direction of the transitions. In our example, the pairs of corresponding states are the two
initial states $(s_{1, 1}, s_{1, 4})$, the two intermediate states $(s_{1, 2}, s_{1, 5})$, and the two final
states $(s_{1, 3}, s_{1, 6})$. We can say that \textit{when moving vertically the current stage of the
replicas is preserved}.

In addition to the exponentially timed $\tau$-transition belonging to the replica, any two states traversed
by the same replica can only possess transitions that are pairwise identically labeled. Those transitions
are originated from (the local states of) sequential processes that are in parallel with (the local state
of) the sequential process originating the considered reducible computation. The set of those transitions
not belonging to the replica can thus be viewed as the \textit{context} of the replica. In our example, the
context of the top replica has a single transition labeled with $\lap a, \lambda \rap$, whereas the context
of the bottom replica is empty. Thus, \textit{when moving horizontally the context of each replica is
preserved}, i.e., the context does not change along a replica. On the other hand, \textit{different replicas
may have different contexts}.

With regard to the identification of the boundary of the replicas of a reducible computation, there are two
possibilities. One is that the final states have no exponentially timed $\tau$-transition, as in our
example. The other is that, at a certain point, each replica has an exponentially timed $\tau$-transition
back to one of the preceding states of the replica itself, as shown below with a variant of our example:
\\[0.1cm]
\centerline{\epsfbox{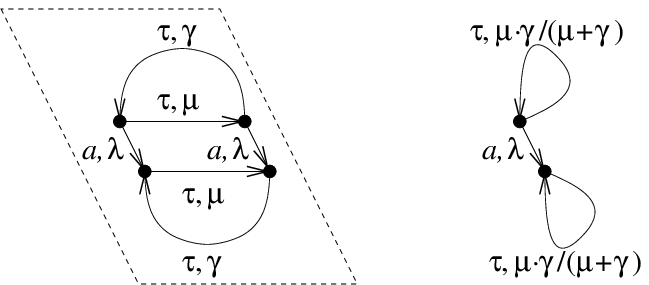}}
In this case, for each replica we view its return state as being its final state. In the figure above, for
both replicas the final state coincides with the initial state.

The new notion of replicated reducible computation must be accompanied by an adjustment of the way measure
$\ms{probtime}$ and multiset $\ms{pbtm}$ are calculated. Given a computation $c$ of the form \linebreak
$P_{1} \arrow{\tau, \lambda_{1}}{\rm M} P_{2} \arrow{\tau, \lambda_{2}}{\rm M} \dots \arrow{\tau,
\lambda_{n}}{\rm M} P_{n + 1}$ that is reducible in the sense of Def.~\ref{reducompdef}, the denominator of
the $i$-th fraction occurring in each of the two factors of $\ms{probtime}(c)$ can indifferently be
$\ms{rate}(P_{i}, \tau, \procs_{\rm M})$ or $\ms{rate}_{\rm t}(P_{i})$: those two values coincide because
$P_{i} \in \procs_{\rm M, fu}$ for all $i = 1, \dots, n$. In contrast, if the reducible computation $c$ is
replicated, each of its replicas has a possibly different context and it is fundamental that
$\ms{rate}(P_{i}, \tau, \procs_{\rm M})$ values are taken as denominators, so as to focus on
$\tau$-transitions. Since there can be $\tau$-transitions also in the context, each destination of those
exit rates needs to be a specific set $\calp$ containing only the states traversed by the replicas rather
than the generic set $\procs_{\rm M}$. Taking into account only $\tau$-transitions leading to states in
$\calp$ ensures \textit{context independence} in this concurrent setting, which opens the way to the
achievement of the same $\ms{probtime}$ value for all the replicas of a reducible computation.

We are by now ready to provide the definition of replicated reducible computation together with the revision
of both $\ms{probtime}$ and $\ms{pbtm}$. Since several reducible computations can depart from the same state
(see the second and the third pair of process terms of Ex.~\ref{wmbeex}), in general we will have to handle
\textit{replicated trees of reducible computations} rather than replicated individual reducible
computations.

In the sequel, we consider $m \in \natns_{> 0}$ process terms $P_{1}, P_{2}, \dots, P_{m} \in \procs_{\rm
M}$ different from each other. We suppose that $P_{k} \arrow{a_{k}, \lambda_{k}}{\rm M} P_{k + 1}$ for all
$k = 1, \dots, m - 1$, with $P_{k}$ having a nonempty tree of computations that are locally reducible for
all $k = 1, \dots, m$ (see $s_{1, 1}$ and $s_{1, 4}$ in our example). This tree is formalized as the set
$C^{\tau}_{k}$ of all the finite-length computations starting from $P_{k}$ such that each of them (i)~is
labeled with a sequence of exponentially timed $\tau$-actions, (ii)~traverses states that are all different
with the possible exception of the final state and one of its preceding states, and (iii)~shares no
transitions with computations in $C^{\tau}_{k'}$ for all $k' \neq k$.

We further suppose that the union of $C^{\tau}_{1}, C^{\tau}_{2}, \dots, C^{\tau}_{m}$ can be partitioned
into $n \in \natns_{> 0}$ \textit{groups of replicas} each consisting of $m$ computations from all the
$m$~sets, such that all the computations in the same group have the same length and are labeled with the
same sequence of exponentially timed $\tau$-actions. As a consequence, for all $k = 1, \dots, m$ we can
write:
\cws{0}{C^{\tau}_{k} \: = \: \{ c_{k, i} \equiv P_{k, i, 1} \arrow{\tau, \lambda_{i, 1}}{\rm M} P_{k, i, 2}
\arrow{\tau, \lambda_{i, 2}}{\rm M} \dots \arrow{\tau, \lambda_{i, l_{i}}}{\rm M} P_{k, i, l_{i} + 1} \mid 1
\le i \le n \}}
where $P_{k, i, 1} \equiv P_{k}$ is the initial state and $l_{i} \in \natns_{> 0}$ is the length of the
computation for all $i = 1, \dots, n$.

	\begin{definition}\label{greducompdef}

The family of computations $\calc^{\tau} = \{ C^{\tau}_{1}, C^{\tau}_{2}, \dots, C^{\tau}_{m} \}$ is said to
be generally reducible, or g-reducible for short, iff \underline{either} $m = 1$ and for all $i = 1, \dots,
n$:

		\begin{itemize}

\item $P_{1, i, j} \in \procs_{\rm M, fu}$ for all $j = 1, \dots, l_{i}$;

\item $P_{1, i, l_{i} + 1} \in \procs_{\rm M, nfu}$ or $P_{1, i, l_{i} + 1} \equiv P_{1, i, j}$ for some $j
= 1, \dots, l_{i}$;

		\end{itemize}

\noindent
\underline{or} $m \ge 1$, with $P_{1, i, j} \in \procs_{\rm M, nfu}$ for all $i = 1, \dots, n$ and $j = 1,
\dots, l_{i}$ when $m = 1$, and for all $i = 1, \dots, n$:

		\begin{itemize}

\item For all $k = 1, \dots, m$, $j = 1, \dots, l_{i}$, and $\lap a, \lambda \rap \in \ms{Act}_{\rm M}$:

			\begin{enumerate}

\item \mbox{\rm [Deviation from the replica]} If $P_{k, i, j} \arrow{a, \lambda}{\rm M} P'$ with $P'
\not\equiv P_{k, i, j + 1}$, then:

				\begin{enumerate}

\item[a.] \mbox{\rm [change of replica via context]} either $P' \equiv P_{k', i, j}$ for some $k' = 1,
\dots, m$;

\item[b.] \mbox{\rm [change of computation]} or $P' \equiv P_{k, i', j'}$ with $a = \tau$ and $\lambda =
\lambda_{i', j' - 1}$ for some $i' = 1, \dots, n$ other than $i$ and some $j' = 2, \dots, l_{i' + 1}$.

				\end{enumerate}

\item \mbox{\rm [Context preservation along the replica]} For all $k' = 1, \dots, m$, it holds that $P_{k,
i, j} \arrow{a, \lambda}{\rm M} P_{k', i, j}$ iff $P_{k, i, j'} \arrow{a, \lambda}{\rm M} P_{k', i, j'}$ for
all $j' = 1, \dots, l_{i}$.

\item \mbox{\rm [Stage preservation across replicas]} For all $i' = 1, \dots, n$ other than $i$ and $j' = 2,
\dots, l_{i' + 1}$, \linebreak it holds that $P_{k, i, j} \arrow{a, \lambda}{\rm M} P_{k, i', j'}$ iff
$P_{k', i, j} \arrow{a, \lambda}{\rm M} P_{k', i', j'}$ for all $k' = 1, \dots, m$.

			\end{enumerate}

\item \mbox{\rm [Termination]} One of the following holds:

			\begin{enumerate}

\item[$\overline{\it 4}$.] Whenever there exists $\lambda_{i, l_{i} + 1} \in \realns_{> 0}$ such that $P_{k,
i, l_{i} + 1} \arrow{\tau, \lambda_{i, l_{i} + 1}}{\rm M} P_{k, i, l_{i} + 2}$ for all $k = 1, \dots, m$,
then at least one of conditions~1, 2, and~3 above is not satisfied by $P_{k', i, l_{i} + 1}$ for some $k' =
1, \dots, m$.

\item[$\widetilde{\it 4}$.] There is no $\lambda_{i, l_{i} + 1} \in \realns_{> 0}$ such that $P_{k, i, l_{i}
+ 1} \arrow{\tau, \lambda_{i, l_{i} + 1}}{\rm M} P_{k, i, l_{i} + 2}$ for all $k = 1, \dots, m$.

\item[$\widehat{\it 4}$.] $P_{k, i, l_{i} + 1} \equiv P_{k, i, j}$ for all $k = 1, \dots, m$ and some $j =
1, \dots, l_{i}$.
\fullbox

			\end{enumerate}

		\end{itemize}

	\end{definition}


\noindent
Some comments are now in order:

	\begin{itemize}

\item In the case that $m = 1$ and all the traversed states are fully unstable (see the ``either'' option),
Def.~\ref{greducompdef} coincides with Def.~\ref{reducompdef} except for the fact that the former considers
a tree of computations whilst the latter considers a single computation.

\item The case $m = 1$ with $P_{1, i, j} \in \procs_{\rm M, nfu}$ for every $i = 1, \dots, n$ and $j = 1,
\dots, l_{i}$ happens when all the sequential process terms in parallel with the one originating the tree of
locally reducible computations repeatedly execute a single action (selfloop transition), thus causing no
replica of the tree to be formed. Both this case and the case $m \ge 2$ are subject to conditions~1, 2, 3,
and~4.

\item Condition 1 establishes that each transition deviating (see $P' \not\equiv P_{k, i, j + 1}$) from the
replica of the considered computation of $\calc^{\tau}$:

		\begin{itemize}

\item either is a vertical transition of the context that preserves the current stage of the replicas and
hence causes the passage to the corresponding state of another replica ($k' \neq k$) or to the same state of
the same replica ($k' = k$, meaning that one of the sequential process terms in parallel with the one
originating the considered computation repeatedly executes a single action);

\item or is a transition belonging to some other computation in $\calc^{\tau}$ starting from the same
process term $P_{k}$ as the considered computation.

		\end{itemize}

These two facts together imply the maximality of $\calc^{\tau}$, because taking into account deviating
transitions causes all replicas to be included. In addition, they prevent process terms like \linebreak
$\lap \tau, \mu \rap . (\lap \tau, \gamma \rap . \nil + \lap a, \lambda \rap . \nil) + \lap a, \lambda \rap
. \nil$ and $\lap \tau, {\mu \cdot \gamma \over \mu + \gamma} \rap . \nil + \lap a, \lambda \rap . \nil$ --
which do not contain occurrences of parallel composition ($m = 1$) and have no fully unstable states when $a
\neq \tau$ -- from being deemed to be equivalent.

\item Condition 2 is related to condition 1.a and ensures that the context of a replica is preserved along
each state traversed by the replica.

\item Condition 3 is related to condition 1.b and ensures that any transition belonging neither to the
considered computation nor to its context (i.e., belonging to some other computation in $\calc^{\tau}$) is
present at the same stage of each replica of the considered computation.

\item The three variants of condition 4 establish the boundary of the replicas of the considered computation
in a way that guarantees the maximality of the length of the replicas themselves under (i)~conditions~1, 2,
and~3, (ii) the constraint that all of their transitions are labeled with exponentially timed
$\tau$-actions, (iii) and the constraint that all the traversed states are different with the possible
exception of the final state and one of its preceding states.

	\end{itemize}

Let $\ms{initial}(\calc^{\tau}) = \{ P_{k} \mid 1 \le k \le m \}$ and $\ms{final}(\calc^{\tau}) = \{ P_{k,
i, l_{i} + 1} \mid 1 \le k \le m, 1 \le i \le n \}$ be the sets of initial states and final states of the
computations in $\calc^{\tau}$. In order to avoid interferences between the computations in $C^{\tau}_{1},
C^{\tau}_{2}, \dots, C^{\tau}_{m}$ and the transitions belonging to the context of those computations, for
any computation $c_{k, i}$ in $\calc^{\tau}$ we consider the following context-free measure:
\cws{0}{\ms{probtime}_{\rm cf}(c_{k, i}) \: = \: \left( \prod\limits_{j = 1}^{l_{i}} {\lambda_{i, j} \over
\ms{rate}(P_{k, i, j}, \tau, \calp_{k})} \right) \cdot \left( \sum\limits_{j = 1}^{l_{i}} {1 \over
\ms{rate}(P_{k, i, j}, \tau, \calp_{k})} \right)}
where $\calp_{k} = \{ P_{k, i', j'} \mid 1 \le i' \le n, 2 \le j' \le l_{i' + 1} \}$. In this way, all
replicas of the same computation will have the same $\ms{probtime}_{\rm cf}$ measure, as shown below.

	\begin{proposition}\label{probtimecfprop}

Whenever $\calc^{\tau}$ is g-reducible, then for all $k, k' = 1, \dots, m$ and $i = 1, \dots, n$:
\cws{12}{\ms{probtime}_{\rm cf}(c_{k, i}) \: = \: \ms{probtime}_{\rm cf}(c_{k', i})}
\fullbox

	\end{proposition}

Moreover, we replace the generic multiset $\ms{pbtm}(P, D)$ with the more specific multisets $\ms{pbtm}_{\rm
cf}(P_{k}, D \cap \ms{final}(\calc^{\tau}))$ for all $P_{k} \in \ms{initial}(\calc^{\tau})$. The latter
multisets are based on $\ms{probtime}_{\rm cf}$ instead of $\ms{probtime}$ as well as on $\ms{reducomp}_{\rm
cf}$ instead of $\ms{reducomp}$, where $\ms{reducomp}_{\rm cf}(P_{k}, D \cap \ms{final}(\calc^{\tau}), t)$
is the multiset of computations identical to those in $C^{\tau}_{k}$ that go from $P_{k}$ to $D \cap
\ms{final}(\calc^{\tau})$ and have average duration~$t$. We point out that computations of length zero are
not considered as $t \in \realns_{> 0}$, so that whenever $P_{k} \in \ms{initial}(\calc^{\tau}) \cap D \cap
\ms{final}(\calc^{\tau})$, then the calculation of $\ms{pbtm}_{\rm cf}(P_{k}, D \cap
\ms{final}(\calc^{\tau}))$ does take into account computations identical to those in $C^{\tau}_{k}$ going
from $P_{k}$ to itself.

	\begin{proposition}\label{pbtmcfprop}

Whenever $\calc^{\tau}$ is g-reducible, then for all $k, k' = 1, \dots, m$:
\cws{12}{\ms{pbtm}_{\rm cf}(P_{k}, \ms{final}(\calc^{\tau})) \: = \: \ms{pbtm}_{\rm cf}(P_{k'},
\ms{final}(\calc^{\tau}))}
\fullbox

	\end{proposition}

We are finally ready to introduce the revised definition of weak Markovian bisimilarity.

	\begin{definition}

An equivalence relation $\calb$ over $\procs_{\rm M}$ is a g-weak Markovian bisimulation iff, whenever
$(P_{1}, P_{2}) \in \calb$, then:

		\begin{itemize}

\item For all visible action names $a \in \ms{Name}_{\rm v}$ and equivalence classes $D \in \procs_{\rm M} /
\calb$:
\cws{12}{\hspace*{-0.6cm} \ms{rate}(P_{1}, a, D) \: = \: \ms{rate}(P_{2}, a, D)}

\item If $P_{1}$ is not an initial state of any g-reducible family of computations, then $P_{2}$ is not an
initial state of any g-reducible family of computations either, and for all equivalence classes $D \in
\procs_{\rm M} / \calb$:
\cws{12}{\hspace*{-0.6cm} \ms{rate}(P_{1}, \tau, D) \: = \: \ms{rate}(P_{2}, \tau, D)}

\item If $P_{1}$ is an initial state of some g-reducible family of computations, then $P_{2}$ is an initial
state of some g-reducible family of computations too, and for all g-reducible families of computations
$\calc^{\tau}_{1}$ with $P_{1} \in \ms{initial}(\calc^{\tau}_{1})$ there exists a g-reducible family of
computations $\calc^{\tau}_{2}$ with $P_{2} \in \ms{initial}(\calc^{\tau}_{2})$ such that for all
equivalence classes $D \in \procs_{\rm M} / \calb$:
\cws{12}{\hspace*{-0.6cm} \ms{pbtm}_{\rm cf}(P_{1}, D \cap \ms{final}(\calc^{\tau}_{1})) \: = \:
\ms{pbtm}_{\rm cf}(P_{2}, D \cap \ms{final}(\calc^{\tau}_{2}))}

		\end{itemize}

\noindent
G-weak Markovian bisimilarity $\wbis{\rm MB, g}$ is the largest g-weak Markovian bisimulation.
\fullbox

	\end{definition}

	\begin{example}\label{gwmbeex}

The process terms mentioned in each of the three cases of Ex.~\ref{wmbeex} are still related by~$\wbis{\rm
MB, g}$. Note that each of those process terms is the only initial state of a g-reducible family of
computations composed by a single computation (first case) or a single tree of computations (second and
third case) traversing only fully unstable states, thus $m = 1$ and the ``either'' option of
Def.~\ref{greducompdef} applies.
\fullbox

	\end{example}

	\begin{example}\label{gwmbecongrparex}

Let us reconsider the two process terms at the beginning of Ex.~\ref{wmbenocongrparex}. Now we have:
\cws{0}{\lap a, \lambda \rap . \lap \tau, \mu \rap . \lap \tau, \gamma \rap . \nil \: \wbis{\rm MB, g} \:
\lap a, \lambda \rap . \lap \tau, {\mu \cdot \gamma \over \mu + \gamma} \rap . \nil}
and:
\cws{0}{\lap a, \lambda \rap . \lap \tau, \mu \rap . \lap \tau, \gamma \rap . \nil \pco{\emptyset} \lap a',
\lambda' \rap . \nil \: \wbis{\rm MB, g} \: \lap a, \lambda \rap . \lap \tau, {\mu \cdot \gamma \over \mu +
\gamma} \rap . \nil \pco{\emptyset} \lap a', \lambda' \rap . \nil}
because it holds that:
\cws{0}{\lap \tau, \mu \rap . \lap \tau, \gamma \rap . \nil \pco{\emptyset} \lap a', \lambda' \rap . \nil \:
\wbis{\rm MB, g} \: \lap \tau, {\mu \cdot \gamma \over \mu + \gamma} \rap . \nil \pco{\emptyset} \lap a',
\lambda' \rap . \nil}
In fact, for $a' \neq \tau$ the two process terms are the initial states of two g-reducible families of
computations $\calc^{\tau}_{1}$ and $\calc^{\tau}_{2}$, respectively, each composed of two replicas -- the
first one having context $\{ \lap a', \lambda' \rap \}$ and final state $\nil \pco{\emptyset} \lap a',
\lambda' \rap . \nil$ and the second one having empty context and final state $\nil \pco{\emptyset} \nil$ --
with:
\cws{0}{\begin{array}{rcl}
\ms{pbtm}_{\rm cf}(\lap \tau, \mu \rap . \lap \tau, \gamma \rap . \nil \pco{\emptyset} \lap a', \lambda'
\rap . \nil, D \cap \ms{final}(\calc^{\tau}_{1})) & = & \lmp {1 \over \mu} + {1 \over \gamma} \rmp \\
\ms{pbtm}_{\rm cf}(\lap \tau, {\mu \cdot \gamma \over \mu + \gamma} \rap . \nil \pco{\emptyset} \lap a',
\lambda' \rap . \nil, D \cap \ms{final}(\calc^{\tau}_{2})) & = & \lmp {\mu + \gamma \over \mu \cdot \gamma}
\rmp \\
\end{array}}
whenever $D$ contains the final state $\nil \pco{\emptyset} \lap a', \lambda' \rap . \nil$, as the way of
calculating $\ms{probtime}_{\rm cf}$ and $\ms{pbtm}_{\rm cf}$ does not take the context into account. \\
For $a' = \tau$, in addition to $\calc^{\tau}_{1}$ and $\calc^{\tau}_{2}$, the two process terms are the
initial states of two further g-reducible families of computations $\calc'^{\tau}_{1}$ and
$\calc'^{\tau}_{2}$, respectively, each composed of two replicas of length 1 labeled with $\lap a', \lambda'
\rap$. In this case:
\cws{0}{\begin{array}{rcl}
\ms{pbtm}_{\rm cf}(\lap \tau, \mu \rap . \lap \tau, \gamma \rap . \nil \pco{\emptyset} \lap a', \lambda'
\rap . \nil, D \cap \ms{final}(\calc'^{\tau}_{1})) & = & \lmp {1 \over \lambda'} \rmp \\[0.1cm]
\ms{pbtm}_{\rm cf}(\lap \tau, {\mu \cdot \gamma \over \mu + \gamma} \rap . \nil \pco{\emptyset} \lap a',
\lambda' \rap . \nil, D \cap \ms{final}(\calc'^{\tau}_{2})) & = & \lmp {1 \over \lambda'} \rmp \\
\end{array}}
whenever $D$ contains the two $\wbis{\rm MB, g}$-equivalent final states $\lap \tau, \mu \rap . \lap \tau,
\gamma \rap . \nil \pco{\emptyset} \nil$ and $\lap \tau, {\mu \cdot \gamma \over \mu + \gamma} \rap . \nil
\pco{\emptyset} \nil$. \\
The two divergent process terms at the end of Ex.~\ref{wmbenocongrparex} are not related by $\wbis{\rm MB,
g}$ because $\gamma_{1} \neq \gamma_{2}$; hence, they no longer result in a disruption of compositionality
when placed in the context $\_ \pco{\emptyset} \lap a', \lambda' \rap . \nil$.
\fullbox

	\end{example}

We conclude by showing that there exists a relationship between $\wbis{\rm MB, g}$ and $\wbis{\rm MB}$ only
for process terms that have no cycles of exponentially timed $\tau$-actions. The reason of this limitation
is that $\wbis{\rm MB, g}$ imposes checks on those cycles that are not always performed by $\wbis{\rm MB}$,
like, e.g., in the case of the two divergent process terms $\textrm{rec} \, X : \lap \tau, \gamma_{1} \rap .
X$ and $\textrm{rec} \, X : \lap \tau, \gamma_{2} \rap . X$ where $\gamma_{1} \neq \gamma_{2}$.

	\begin{proposition}\label{inclusionprop}

Let $P_{1}, P_{2} \in \procs_{\rm M}$ be not divergent. Then:
\cws{12}{P_{1} \wbis{\rm MB} P_{2} \: \Longrightarrow \: P_{1} \wbis{\rm MB, g} P_{2}}
\fullbox

	\end{proposition}

%
\subsection{Congruence Property}\label{gwmbccongr}
%

The investigation of the compositionality of $\wbis{\rm MB, g}$ with respect to MPC operators leads to
results analogous to those for $\wbis{\rm MB}$ \cite{Ber}, plus the achievement of congruence with respect
to parallel composition.

	\begin{proposition}\label{gwmbecongrprop}

Let $P_{1}, P_{2} \in \procs_{\rm M}$. Whenever $P_{1} \wbis{\rm MB, g} P_{2}$, then:

		\begin{enumerate}

\item $\lap a, \lambda \rap . P_{1} \wbis{\rm MB, g} \lap a, \lambda \rap . P_{2}$ for all $\lap a, \lambda
\rap \in \ms{Act}_{\rm M}$.

\item $P_{1} / H \wbis{\rm MB, g} P_{2} / H$ for all $H \subseteq \ms{Name}_{\rm v}$.

\item $P_{1} \pco{S} P \wbis{\rm MB, g} P_{2} \pco{S} P$ and $P \pco{S} P_{1} \wbis{\rm MB, g} P \pco{S}
P_{2}$ for all $S \subseteq \ms{Name}_{\rm v}$ and $P \in \procs_{\rm M}$.
\fullbox

		\end{enumerate}

	\end{proposition}

The relation $\wbis{\rm MB, g}$ is not a congruence with respect to the alternative composition operator due
to fully unstable process terms: for instance, it holds that $\lap \tau, \mu \rap . \lap \tau, \gamma \rap .
\nil \wbis{\rm MB, g} \lap \tau, {\mu \cdot \gamma \over \mu + \gamma} \rap . \nil$ whereas \linebreak $\lap
\tau, \mu \rap . \lap \tau, \gamma \rap . \nil + \lap a, \lambda \rap . \nil \not\wbis{\rm MB, g} \lap \tau,
{\mu \cdot \gamma \over \mu + \gamma} \rap . \nil + \lap a, \lambda \rap . \nil$. In fact, if it were $a
\neq \tau$, then we would have:
\cws{0}{\begin{array}{rcl}
\ms{rate}(\lap \tau, \mu \rap . \lap \tau, \gamma \rap . \nil + \lap a, \lambda \rap . \nil, \tau,
[\nil]_{\wbis{\rm MB, g}}) & = & 0 \\
\ms{rate}(\lap \tau, {\mu \cdot \gamma \over \mu + \gamma} \rap . \nil + \lap a, \lambda \rap . \nil, \tau,
[\nil]_{\wbis{\rm MB, g}}) & = & {\mu \cdot \gamma \over \mu + \gamma} \\
\end{array}}
otherwise for $a = \tau$ the two process terms would be the initial states of two g-reducible families of
computations, respectively, each composed of a single tree of computations with final state $\nil$ and we
would have:
\cws{6}{\begin{array}{rcl}
\ms{pbtm}_{\rm cf}(\lap \tau, \mu \rap . \lap \tau, \gamma \rap . \nil + \lap a, \lambda \rap . \nil, \{
\nil \}) & = & \lmp {\mu \over \mu + \lambda} \cdot \left( {1 \over \mu + \lambda} + {1 \over \gamma}
\right), {\lambda \over \mu + \lambda} \cdot {1 \over \mu + \lambda} \rmp \\
\ms{pbtm}_{\rm cf}(\lap \tau, {\mu \cdot \gamma \over \mu + \gamma} \rap . \nil + \lap a, \lambda \rap .
\nil, \{ \nil \}) & = & \lmp {1 \over {{\mu \cdot \gamma \over \mu + \gamma} + \lambda}} \rmp \\
\end{array}}

The congruence violation with respect to the alternative composition operator can be prevented by adopting a
construction analogous to the one used in~\cite{Mil} for weak bisimilarity over nondeterministic process
terms and adapted in~\cite{Ber} to $\wbis{\rm MB}$. Therefore, we have to apply the exit rate equality check
for $\tau$-actions also to process terms that are initial states of g-reducible families of computations,
with the equivalence classes to consider being the ones with respect to $\wbis{\rm MB, g}$.

	\begin{definition}

Let $P_{1}, P_{2} \in \procs_{\rm M}$. We say that $P_{1}$ is g-weakly Markovian bisimulation congruent to
$P_{2}$, written $P_{1} \obis{\rm MB, g} P_{2}$, iff for all action names $a \in \ms{Name}$ and equivalence
classes $D \in \procs_{\rm M} / \! \wbis{\rm MB, g}$:
\cws{12}{\ms{rate}(P_{1}, a, D) \: = \: \ms{rate}(P_{2}, a, D)}
\fullbox

	\end{definition}

	\begin{proposition}\label{gbehavequivinclprop}

$\sbis{\rm MB} \, \subset \, \obis{\rm MB, g} \, \subset \, \wbis{\rm MB, g}$, with $\obis{\rm MB, g} \, =
\, \wbis{\rm MB, g}$ over the set of process terms of $\procs_{\rm M}$ that are not initial states of any
g-reducible family of computations.
\fullbox

	\end{proposition}

	\begin{proposition}\label{gwmbewmbcprop}

Let $P_{1}, P_{2} \in \procs_{\rm M}$ and $\lap a, \lambda \rap \in \ms{Act}_{\rm M}$. Then:
\cws{12}{\lap a, \lambda \rap . P_{1} \obis{\rm MB, g} \lap a, \lambda \rap . P_{2} \: \Longleftrightarrow
\: P_{1} \wbis{\rm MB, g} P_{2}}
\fullbox

	\end{proposition}

The relation $\obis{\rm MB, g}$ turns out to be the coarsest congruence -- with respect to all the operators
of MPC as well as recursion -- contained in $\wbis{\rm MB, g}$, as shown below.

	\begin{theorem}\label{gwmbccongrthm}

Let $P_{1}, P_{2} \in \procs_{\rm M}$. Whenever $P_{1} \obis{\rm MB, g} P_{2}$, then:

		\begin{enumerate}

\item $\lap a, \lambda \rap . P_{1} \obis{\rm MB, g} \lap a, \lambda \rap . P_{2}$ for all $\lap a, \lambda
\rap \in \ms{Act}_{\rm M}$.

\item $P_{1} + P \obis{\rm MB, g} P_{2} + P$ and $P + P_{1} \obis{\rm MB, g} P + P_{2}$ for all $P \in
\procs_{\rm M}$.

\item $P_{1} / H \obis{\rm MB, g} P_{2} / H$ for all $H \subseteq \ms{Name}_{\rm v}$.

\item $P_{1} \pco{S} P \obis{\rm MB, g} P_{2} \pco{S} P$ and $P \pco{S} P_{1} \obis{\rm MB, g} P \pco{S}
P_{2}$ for all $S \subseteq \ms{Name}_{\rm v}$ and $P \in \procs_{\rm M}$.
\fullbox

		\end{enumerate}

	\end{theorem}

	\begin{theorem}\label{gwmbccoarsestcongrthm}

Let $P_{1}, P_{2} \in \procs_{\rm M}$. Then $P_{1} \obis{\rm MB, g} P_{2}$ iff $P_{1} + P \wbis{\rm MB, g}
P_{2} + P$ for all $P \in \procs_{\rm M}$.
\fullbox

	\end{theorem}

With regard to recursion, we need to extend $\obis{\rm MB, g}$ to open process terms in the usual way.
Similar to other congruence proofs for bisimulation equivalence with respect to recursion, here we rely on a
notion of g-weak Markovian bisimulation up to $\wbis{\rm MB, g}$ inspired by the notion of Markovian
bisimulation up to $\sbis{\rm MB}$ of~\cite{BBG}. This notion differs from its nondeterministic counterpart
used in~\cite{Mil} due to the necessity of working with equivalence classes in this Markovian setting.

	\begin{definition}

Let $P_{1}, P_{2} \in \calpl_{\rm M}$ be process terms containing free occurrences of $k \in \natns$ process
variables $X_{1}, \ldots, X_{k} \in \ms{Var}$ at most. We define $P_{1} \obis{\rm MB, g} P_{2}$ iff $P_{1}
\{ Q_{i} \hookrightarrow X_{i} \mid 1 \le i \le k \} \obis{\rm MB, g} P_{2} \{ Q_{i} \hookrightarrow X_{i}
\mid 1 \le i \le k \}$ for all $Q_{1}, \dots, Q_{k} \in \calpl_{\rm M}$ containing no free occurrences of
process variables.
\fullbox

	\end{definition}

	\begin{definition}

Let $^{+}$ denote the operation of transitive closure for relations. A binary relation $\calb$
over~$\procs_{\rm M}$ is a g-weak Markovian bisimulation up to $\wbis{\rm MB, g}$ iff, whenever $(P_{1},
P_{2}) \in \calb$, then:

		\begin{itemize}

\item For all visible action names $a \in \ms{Name}_{\rm v}$ and equivalence classes $D \in \procs_{\rm M} /
(\calb \cup \calb^{-1} \cup \wbis{\rm MB, g})^{+}$:
\cws{12}{\hspace*{-0.6cm} \ms{rate}(P_{1}, a, D) \: = \: \ms{rate}(P_{2}, a, D)}

\item If $P_{1}$ is not an initial state of any g-reducible family of computations, then $P_{2}$ is not
\linebreak an initial state of any g-reducible family of computations either, and for all equivalence
classes $D \in \procs_{\rm M} / (\calb \cup \calb^{-1} \cup \wbis{\rm MB, g})^{+}$:
\cws{12}{\hspace*{-0.6cm} \ms{rate}(P_{1}, \tau, D) \: = \: \ms{rate}(P_{2}, \tau, D)}

\item If $P_{1}$ is an initial state of some g-reducible family of computations, then $P_{2}$ is an initial
state of some g-reducible family of computations too, and for all g-reducible families of computations
$\calc^{\tau}_{1}$ with $P_{1} \in \ms{initial}(\calc^{\tau}_{1})$ there exists a g-reducible family of
computations $\calc^{\tau}_{2}$ with $P_{2} \in \ms{initial}(\calc^{\tau}_{2})$ such that for all
equivalence classes $D \in \procs_{\rm M} / (\calb \cup \calb^{-1} \cup \wbis{\rm MB, g})^{+}$:
\cws{12}{\hspace*{-0.6cm} \ms{pbtm}_{\rm cf}(P_{1}, D \cap \ms{final}(\calc^{\tau}_{1})) \: = \:
\ms{pbtm}_{\rm cf}(P_{2}, D \cap \ms{final}(\calc^{\tau}_{2}))}
\fullbox

		\end{itemize}

	\end{definition}

	\begin{proposition}\label{gwmbeuptoprop}

Let $\calb$ be a relation over $\procs_{\rm M}$. If $\calb$ is a g-weak Markovian bisimulation up to
$\wbis{\rm MB, g}$, then $(P_{1}, P_{2}) \in \calb$ implies $P_{1} \wbis{\rm MB, g} P_{2}$ for all $P_{1},
P_{2} \in \procs_{\rm M}$. Moreover $(\calb \cup \calb^{-1} \cup \wbis{\rm MB, g})^{+} = \; \wbis{\rm MB,
g}$.
\fullbox

	\end{proposition}

	\begin{theorem}\label{gwmbccongrrecthm}

Let $P_{1}, P_{2} \in \calpl_{\rm M}$ be process terms containing free occurrences of $k \in \natns$ process
variables $X_{1}, \ldots, X_{k} \in \ms{Var}$ at most. Whenever $P_{1} \obis{\rm MB, g} P_{2}$, then:
\cws{12}{\textrm{rec} \, X_{1} : \dots : \textrm{rec} \, X_{k} : P_{1} \: \obis{\rm MB, g} \: \textrm{rec}
\, X_{1} : \dots : \textrm{rec} \, X_{k} : P_{2}}
\fullbox

	\end{theorem}

%
\subsection{Exactness at Steady State}\label{gwmbcexact}
%

We conclude by examining the exactness of the CTMC-level aggregation induced by $\wbis{\rm MB, g}$ and
$\obis{\rm MB, g}$. In general, a CTMC aggregation is said to be exact at steady state (resp.\ transient
state) iff the steady-state (resp.\ transient) probability of being in a macrostate of an aggregated CTMC is
the sum of the steady-state (resp.\ transient) probabilities of being in each of the constituent microstates
of the original CTMC from which the aggregated one has been obtained. This property implies the preservation
of steady-state (resp.\ transient) reward-based performance measures across CTMC models.

The aggregation to examine -- which we call GW-lumpability -- shares with the one induced by $\wbis{\rm MB}$
and $\obis{\rm MB}$ -- called W-lumpability in~\cite{Ber} -- the characteristic of viewing certain sequences
of exponentially timed $\tau$-actions to be equivalent to individual exponentially timed $\tau$-actions
having the same average duration and the same execution probability as the corresponding sequences when the
latter are considered locally to the processes originating them.\footnote{To be precise, since the Markov
property of the original CTMC is not preserved but the aggregated stochastic process is still assumed to be
a CTMC, it would be more appropriate to call those aggregations pseudo-aggregations~\cite{RS}.} On the other
hand, due to the idea of context embodied in the notion of g-reducible family of computations and the
consequent capability of distinguishing between action disabling and action interruption, a notable
difference between GW-lumpability and W-lumpability is that the former may aggregate states also in the case
of concurrent processes, while the latter cannot.

Reducing a computation formed by at least two exponentially timed $\tau$-transitions to a single
exponentially timed $\tau$-transition with the same average duration amounts to approximating a
hypoexponentially (or Erlang) distributed random variable with an exponentially distributed random variable
having the same expected value. This implies that, in general, GW-lumpability cannot preserve transient
performance measures, as was the case with W-lumpability~\cite{Ber}. However, while W-lumpability at least
preserves transient properties expressed in terms of the mean time to certain events, this is no longer the
case with GW-lumpability as we have seen at the beginning of Sect.~\ref{wmberedef}.

What turns out for GW-lumpability is that, similar to W-lumpability, it preserves steady-state performance
measures, provided that the states traversed by any replica of a reducible computation have the same rewards
and the transitions -- belonging to the replica or to the context -- departing from any two traversed states
have pairwise identical rewards. However, unlike W-lumpability, we have to confine ourselves to processes in
which synchronizations (if any) do not take place right before the beginning of computations that are
reducible according to the ``or'' option of Def.~\ref{greducompdef}. This constraint comes from the
insensitivity conditions for generalized semi-Markov processes mentioned in~\cite{Mat,HL,Hil}.

	\begin{theorem}\label{gwmbexactthm}

GW-lumpability is exact at steady state over every process term $P \in \procs_{\rm M}$ such that, for all
g-reducible families of computations $\calc^{\tau}$ in $\lsp P \rsp_{\rm M}$ with size $m \ge 2$, or size $m
= 1$ and all the traversed states being not fully unstable, no state in $\ms{initial}(\calc^{\tau})$ is the
target state of a transition in $\lsp P \rsp_{\rm M}$ arising from the synchronization of two or more
actions.
\fullbox

	\end{theorem}

	\begin{example}

In order to illustrate the need for the constraint on synchronizations in Thm.~\ref{gwmbexactthm}, consider
the following two process terms:
\cws{0}{\begin{array}{rcl}
P_{1} & \equiv & \textrm{rec} \, X : \lap \tau, \mu \rap . \lap \tau, \gamma \rap . \lap b, \delta \rap . X
\pco{\{ b \}} \textrm{rec} \, Y : \lap a, \lambda \rap . \lap b, \delta \rap . Y \\
P_{2} & \equiv & \textrm{rec} \, X : \lap \tau, {\mu \cdot \gamma \over \mu + \gamma} \rap . \lap b, \delta
\rap . X \pco{\{ b \}} \textrm{rec} \, Y : \lap a, \lambda \rap . \lap b, \delta \rap . Y \\
\end{array}}
Observe that $P_{1} \wbis{\rm MB, g} P_{2}$ and that $\lsp P_{1} \rsp_{\rm M}$ and $\lsp P_{2} \rsp_{\rm M}$
are given by the two labeled multitransition systems depicted at the beginning of Sect.~\ref{wmberedef},
respectively, with an additional transition labeled with $\lap b, \delta \rap$ from the final state to the
initial one. In the case that $\mu = \gamma = \lambda = \delta = 1$ and $\delta \otimes \delta = \delta$, it
turns out that the steady-state probability distribution for $\lsp P_{1} \rsp_{\rm M}$ is as follows:
\cws{0}{\begin{array}{rclcrclcrcl}
\pi[s_{1, 1}] & = & {2 \over 13} & \qquad &
\pi[s_{1, 2}] & = & {1 \over 13} & \qquad &
\pi[s_{1, 3}] & = & {1 \over 13} \\[0.1cm]
\pi[s_{1, 4}] & = & {2 \over 13} & &
\pi[s_{1, 5}] & = & {3 \over 13} & &
\pi[s_{1, 6}] & = & {4 \over 13} \\[0.1cm]
\end{array}}
whereas the steady-state probability distribution for $\lsp P_{2} \rsp_{\rm M}$ is as follows:
\cws{0}{\begin{array}{rclcrcl}
\pi[s_{2, 1}] & = & {2 \over 10} & \qquad &
\pi[s_{2, 2}] & = & {1 \over 10} \\[0.1cm]
\pi[s_{2, 3}] & = & {4 \over 10} & &
\pi[s_{2, 4}] & = & {3 \over 10} \\[0.1cm]
\end{array}}
Thus, the CTMC underlying $\lsp P_{2} \rsp_{\rm M}$ is not an exact aggregation of the CTMC underlying $\lsp
P_{1} \rsp_{\rm M}$ because:
\cws{0}{\begin{array}{rclcrcl}
\pi[s_{1, 1}] + \pi[s_{1, 2}] & \neq & \pi[s_{2, 1}] & \qquad &
\pi[s_{1, 3}] & \neq & \pi[s_{2, 2}] \\[0.1cm]
\pi[s_{1, 4}] + \pi[s_{1, 5}] & \neq & \pi[s_{2, 3}] & &
\pi[s_{1, 6}] & \neq & \pi[s_{2, 4}] \\[0.1cm]
\end{array}}
As can be noted, the transition in $\lsp P_{1} \rsp_{\rm M}$ labeled with $\lap b, \delta \rap$ arises from
the synchronization of two $b$-actions and its target state is the initial state of a computation belonging
to a g-reducible family with size $m = 2$; hence, Thm.~\ref{gwmbexactthm} does not apply.

In contrast, if we consider a synchronization-free variant of the two process terms above like for instance:
\cws{0}{\begin{array}{rcl}
P_{3} & \equiv & \textrm{rec} \, X : \lap \tau, \mu \rap . \lap \tau, \gamma \rap . \lap b_{1}, \delta_{1}
\rap . X \pco{\emptyset} \textrm{rec} \, Y : \lap a, \lambda \rap . \lap b_{2}, \delta_{2} \rap . Y \\
P_{4} & \equiv & \textrm{rec} \, X : \lap \tau, {\mu \cdot \gamma \over \mu + \gamma} \rap . \lap b_{1},
\delta_{1} \rap . X \pco{\emptyset} \textrm{rec} \, Y : \lap a, \lambda \rap . \lap b_{2}, \delta_{2} \rap .
Y \\
\end{array}}
we have that for $\mu = \gamma = \lambda = \delta_{1} = \delta_{2} = 1$ the steady-state probability
distribution for $\lsp P_{3} \rsp_{\rm M}$ is:
\cws{0}{\begin{array}{rclcrclcrcl}
\pi[s_{3, 1}] & = & {1 \over 6} & \qquad &
\pi[s_{3, 2}] & = & {1 \over 6} & \qquad &
\pi[s_{3, 3}] & = & {1 \over 6} \\[0.1cm]
\pi[s_{3, 4}] & = & {1 \over 6} & &
\pi[s_{3, 5}] & = & {1 \over 6} & &
\pi[s_{3, 6}] & = & {1 \over 6} \\[0.1cm]
\end{array}}
and the steady-state probability distribution for $\lsp P_{4} \rsp_{\rm M}$ is:
\cws{0}{\begin{array}{rclcrcl}
\pi[s_{4, 1}] & = & {2 \over 6} & \qquad &
\pi[s_{4, 2}] & = & {1 \over 6} \\[0.1cm]
\pi[s_{4, 3}] & = & {2 \over 6} & &
\pi[s_{4, 4}] & = & {1 \over 6} \\[0.1cm]
\end{array}}
hence the CTMC underlying $\lsp P_{4} \rsp_{\rm M}$ is an exact aggregation of the CTMC underlying $\lsp
P_{3} \rsp_{\rm M}$ because:
\cws{12}{\begin{array}{rclcrcl}
\pi[s_{3, 1}] + \pi[s_{3, 2}] & = & \pi[s_{4, 1}] & \qquad &
\pi[s_{3, 3}] & = & \pi[s_{4, 2}] \\[0.1cm]
\pi[s_{3, 4}] + \pi[s_{3, 5}] & = & \pi[s_{4, 3}] & &
\pi[s_{3, 6}] & = & \pi[s_{4, 4}] \\[0.1cm]
\end{array}}
\fullbox

	\end{example}

%
%
\section{Conclusion}\label{concl}
%
%

In this paper, we have introduced $\wbis{\rm MB, g}$ and $\obis{\rm MB, g}$ as variants of the weak
Markovian bisimulation equivalences $\wbis{\rm MB}$ and $\obis{\rm MB}$ proposed in~\cite{Ber}, which suffer
from a limited usefulness for state space reduction purposes as they are not congruences with respect to the
parallel composition operator. The motivation behind $\wbis{\rm MB, g}$ and $\obis{\rm MB, g}$ is thus that
of retrieving full compositionality. Taking inspiration from the idea of preserving the context
of~\cite{Hil}, this has been achieved by enhancing the abstraction capability -- with respect to $\wbis{\rm
MB}$ and $\obis{\rm MB}$ -- when dealing with concurrent computations. The price to pay for the resulting
compositional abstraction capability is that the exactness at steady state of the induced CTMC-level
aggregation does not hold for all the considered processes -- as it was for $\wbis{\rm MB}$ and $\obis{\rm
MB}$ -- but only for sequential processes with abstraction and concurrent processes whose synchronizations
do not take place right before the beginning of computations to be reduced. Additionally, not even transient
properties expressed in terms of the mean time to certain events are preserved in general.

With regard to~\cite{Hil}, where weak isomorphism has been studied, our equivalences $\wbis{\rm MB, g}$ and
$\obis{\rm MB, g}$ have been developed in the more liberal bisimulation framework. A more important novelty
with respect to weak isomorphism is that we have considered not only individual sequences of exponentially
timed $\tau$-actions. In fact, we have addressed trees of exponentially timed $\tau$-actions and we have
established the conditions under which such trees can be reduced -- also in the presence of parallel
composition -- by locally preserving both the average duration and the execution probability of their
branches.

Another approach to abstracting from $\tau$-actions in an exponentially timed setting comes from~\cite{Bra},
where a variant of Markovian bisimilarity was defined that checks for exit rate equality with respect to all
equivalence classes apart from the one including the processes under examination. Congruence and
axiomatization results were provided for the proposed equivalence, and a logical characterization based on
CSL was illustrated in~\cite{BKHW}. However, unlike $\wbis{\rm MB, g}$ and $\obis{\rm MB, g}$, nothing was
said about exactness.

As far as future work is concerned, we would like to investigate equational and logical characterizations of
$\obis{\rm MB, g}$ as well as conduct case studies for assessing its usefulness in practice (especially with
respect to the constraint on synchronizations that guarantees steady-state exactness). With regard to
verification issues, since $\obis{\rm MB} \, \subset \, \obis{\rm MB, g}$ for non-divergent process terms,
we have that the equivalence checking algorithm developed for $\obis{\rm MB}$ in~\cite{Ber} can be exploited
for compositional state space reduction with respect to $\obis{\rm MB, g}$, by applying it to each of the
sequential processes composed in parallel.

\bigskip
\noindent
\textbf{Acknowledgment}: This work has been funded by MIUR-PRIN project \textit{PaCo -- Performability-Aware
Computing: Logics, Models, and Languages}.

\bibliographystyle{eptcs}

\end{document}

%% file: Commands.tex
\def\ms#1{\null\ifmmode\mathord{\mathcode`-="702D\it #1\mathcode`\-="2200}%
	\else$\mathord{\mathcode`-="702D\it #1\mathcode`\-="2200}$\fi}

\newcommand{\cws}[2]
	{\\ \centerline{$#2$} \\[-#1pt]}

\newlength{\spacelen}

\newcommand{\lap}
	{\mbox{$<$}}
\newcommand{\rap}
	{\mbox{$>$}}

\newcommand{\lsp}
	{[ \! [}
\newcommand{\rsp}
	{] \! ]}

\newcommand{\lmp}
	{\{ \! | \,}
\newcommand{\rmp}
	{\, | \! \}}

\newcommand{\cala}
        {\mathcal{A}}

\newcommand{\calb}
        {\mathcal{B}}

\newcommand{\calc}
        {\mathcal{C}}

\newcommand{\calp}
        {\mathcal{P}}

\newcommand{\calpl}
        {\mathcal{PL}}

\newcommand{\natns}
	{\mathbb{N}}

\newcommand{\realns}
	{\mathbb{R}}

\newcommand{\procs}
	{\mathbb{P}}

\newcommand{\arrow}[2]
        {\, {\auxarrow\limits^{#1}}_{#2} \,}
\newcommand{\auxarrow}
	{\mathop{- \!\! - \!\!\!\! \longrightarrow}}

\newcommand{\nil}
	{\underline 0}

\newcommand{\sbis}[1]
	{\sim_{#1}}

\newcommand{\wbis}[1]
        {\approx_{#1}}

\newcommand{\obis}[1]
        {\simeq_{#1}}

\newcommand{\pco}[1]
	{\mathop{\Vert_{#1}}}

\newcommand{\infr}[2]
	{\renewcommand{\arraystretch}{1.5}
	\begin{array}{c}
	#1\\
	\hline
	#2
	\end{array}}

\newcommand{\fullbox}
	{{\mbox{}\nolinebreak\hfill{$\rule{2mm}{2mm}$}}}



%% file: Environments.tex
\newtheorem{new_theorem}
	{Theorem}[section]

\newtheorem{new_definition}
	[new_theorem]{Definition}

\newtheorem{new_remark}
	[new_theorem]{Remark}

\newtheorem{new_example}
	[new_theorem]{Example}

\newtheorem{new_lemma}
	[new_theorem]{Lemma}

\newtheorem{new_proposition}
	[new_theorem]{Proposition}

\newtheorem{new_corollary}
	[new_theorem]{Corollary}

\newenvironment{definition}
	{\begin{new_definition}\rm}
	{\end{new_definition}}

\newenvironment{example}
	{\begin{new_example}\rm}
	{\end{new_example}}

\newenvironment{proposition}
	{\begin{new_proposition}\rm}
	{\end{new_proposition}}

\newenvironment{theorem}
	{\begin{new_theorem}\rm}
	{\end{new_theorem}}

